\documentclass[aps,prl,twocolumn,superscriptaddress, floatfix]{revtex4-2}

\usepackage{gensymb}  
\usepackage{graphicx}  
\usepackage{hyperref}
\usepackage[table]{xcolor}

\hypersetup{colorlinks=true,urlcolor= blue,citecolor=blue,linkcolor= blue}

\usepackage{enumitem}
\setlist{noitemsep,leftmargin=*,topsep=0pt,parsep=0pt}
\usepackage{chemmacros}
\usepackage{upgreek}

\begin{document}


\title{Temperature Evolution of Domains and Intradomain Chirality in 1T-TaS$_{2}$}


\author{Boning Yu}
\author{Ghilles Ainouche}
\author{Manoj Singh}
\author{Bishnu Sharma}
\author{James Huber}
\author{Michael C. Boyer}
\email{To whom correspondence should be addressed. mboyer@clarku.edu}
\affiliation{Department of Physics, Clark University, Worcester, Massachusetts 01610, USA}

\date{\today}

\begin{abstract}
We use scanning tunneling microscopy to study the temperature evolution of the atomic-scale properties of the nearly-commensurate charge density wave (NC-CDW) state of the low-dimensional material, 1T-TaS$_2$. Our measurements at 203 K, 300 K, and 354 K, roughly spanning the temperature range of the NC-CDW state, show that while the average CDW periodicity is temperature independent, domaining and the local evolution of the CDW lattice within a domain is temperature dependent. Further, we characterize the temperature evolution of the displacement field associated with the recently-discovered intradomain chirality of the NC-CDW state by calculating the local rotation vector. Intradomain chirality throughout the NC-CDW phase is likely driven by a strong coupling of the CDW lattice to the atomic lattice.
\end{abstract}

\maketitle

\section{}

\subsection{Introduction}
1T-TaS$_{2}$, is a member of the transition-metal dichalcogenides (TMDs) and has garnered significant interest for the wealth of physics it hosts including multiple charge density wave (CDW) states  \cite{Tanda1984, Thomson1994}, a reported Mott insulating state \cite{Law2017}, and superconductivity \cite{Sipos2008, Dong2021}. Additional attention has focused on manipulating these states through a variety of means including elemental doping, pressure, thickness, strain, as well as local voltage and light pulses so as to induce transitions from one quantum state to another or to create new orders.\cite{Li2012, Liu2013, Ritschel2013,Yoshida2014, Bu2019, Ma2016, Stojchevska2014} In this work, we focus on one of the CDW states hosted by 1T-TaS$_{2}$, the nearly-commensurate (NC-CDW) phase.

Upon cooling, 1T-TaS$_{2}$ hosts three CDW states including an incommensurate (IC) phase above $\sim$355 K, a nearly-commensurate (NC) phase between $\sim$180 K and $\sim$355 K, and a commensurate (C) phase below $\sim$180 K.\cite{Thomson1994} Within the IC-CDW state, the CDW lattice is hexagonal with a periodicity of 3.55$a_{0}$ and zero rotation relative to the atomic lattice.\cite{Nakanishi1977} In the low-temperature C-CDW state, the CDW lattice is a commensurate $\sqrt{13}$a$_0$$\times$$\sqrt{13}$a$_0$ supermodulation which is rotated 13.9{\degree} relative to the atomic lattice.\cite{Ishiguro1991}

 The NC-CDW phase is a bridge between the IC- and C-CDW states and has an \emph{average} CDW lattice periodicity of $\sim$11.75 \text{\AA} and \emph{average} CDW lattice rotation of $\sim$11-13{\degree}, depending on the temperature, relative to the atomic lattice.\cite{Ishiguro1991, Thomson1994} The NC-CDW state is typically described as comprised of hexagonally-ordered domains within which there is a C-CDW state separated by domain walls characterized by discommensurations or an incommensurate CDW network.\cite{Yoshida2014, Tsen2015, Yoshida2015, Shao2016, Zong2018, Wang2019} Such typical descriptions imply that the CDW lattice within a domain in the NC-CDW state is equivalent to that of the low temperature C-CDW state. However, recent work at 300 K \cite{Singh2022} indicates that the CDW lattice within domains is not equivalent to that of the low-temperature C-CDW state. Rather, instead of a uniform C-CDW state within a domain, there is a continual evolution of the CDW lattice from domain center to domain wall. The CDW lattice parameter and angle relative to the atomic lattice at a domain center is close to that of the low-temperature C-CDW state, but evolves toward the domain wall. One goal of the current work is to understand the CDW lattice within domain and domain walls particularly near the temperature limits of the NC-CDW state.

Making the NC-CDW state of even further interest, the NC-CDW state has recently been reported to have a chiral CDW domain structure.\cite{Singh2022} A select few chiral CDW states among TMDs have been previously reported including in 1T-TiSe$_{2}$ \cite{Ishioka2010} as well as for a specific Ti-doping of 1T-TaS$_{2}$, \cite{Gao2021}. In these compounds, the chiral CDW states are believed to originate from orbital ordering \cite{Gao2021, vanWezel2011}. Within undoped 1T-TaS$_{2}$, the NC-CDW state has one chirality which is a simple in-plane mirror symmetry breaking which occurs when the CDW lattice rotates relative to the atomic lattice on cooling from the IC to the NC-CDW phase.\cite{Zong2018} More recently, measurements at 300 K show that the NC-CDW state has a second chirality of separate origin, an intradomain chirality, which is suggested to originate from the strong coupling of an incommensurate CDW state to the atomic lattice.\cite{Singh2022} A second goal of the current work is to extend studies of this intradomain chirality to the temperature limits of NC-CDW state.

\subsection{CDW Lattice Domain Evolution}
\begin{figure}[h]
\includegraphics[clip=true,width=\columnwidth]{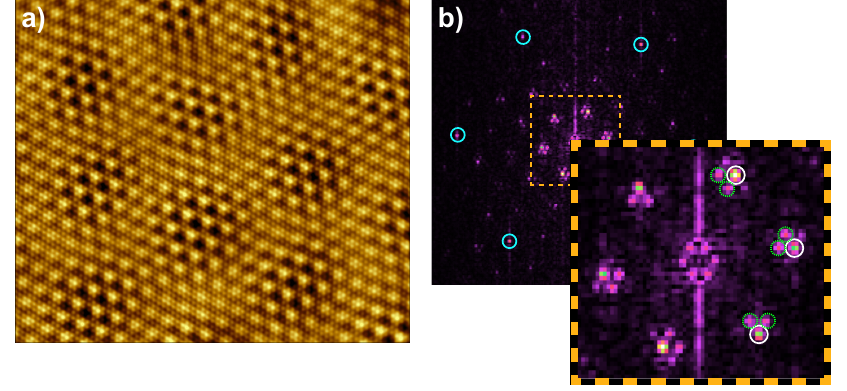}
\caption{a) 230 {\AA} x 200 {\AA} topographic image of 1T-TaS$_2$ Fourier filtered to include all fundamental
atomic, CDW, and domain-related signals. I =  450 pA and V$_{Sample}$ = +150 mV. b) FFT of topography. Atomic periodicities are circled in cyan, central CDW periodicities in white, and satellite peaks in green. Figures in a) and b) are from \cite{Singh2022}}
\label{fig:Fig1}
\end{figure}

In our previous work we detailed the atomic-scale characteristics of the CDW lattice in the NC-CDW state at 300 K.\cite{Singh2022} Here we build upon this initial work and extend it to near the temperature limits of the NC-CDW phase so as to detail and understand the evolution of the CDW lattice throughout this phase. Figure \ref{fig:Fig1}a shows a topographic image of the NC-CDW state we acquired at 300 K. Atomic and CDW periodicities as well as domaining are clearly visible by eye. The wavevectors associated with each of the periodicities can be extracted from the FFT of the topography (Figure \ref{fig:Fig1}b).

\begin{figure}[h]
\includegraphics[clip=true,width=\columnwidth]{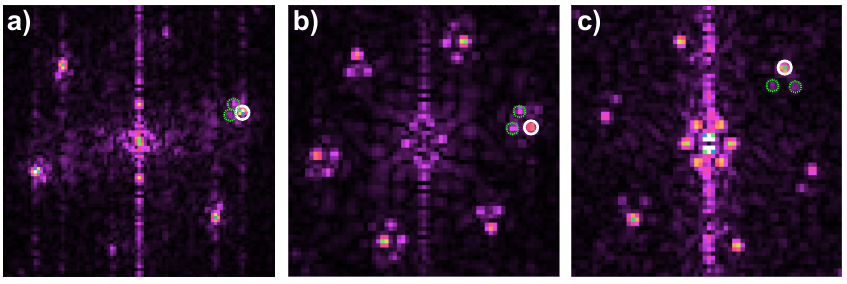}
\caption{FFTs of STM topographies acquired at a) 203 K, b) 300 K, and c) 354 K. In each, a central CDW peak is circled in white, and satellite peaks in green. Note, the FFTs are cropped such that atomic peaks are not included. This allows the details associated with the central CDW signals and surrounding satellite peaks to be clearly seen.}
\label{fig:Fig2}
\end{figure}

Figure \ref{fig:Fig2} shows the FFTs, zoomed to include only CDW-related signals, of acquired STM topographic images taken at 203 K, 300 K, and 354 K. These three temperatures span the majority of the NC-CDW temperature range ($\sim$180 K to $\sim$355 K). In each FFT, circled in white, is a central CDW peak around which there are satellite peaks (circled in green), standard for the NC-CDW state. Most obvious in the FFTs are the change in the distance of the satellite peak to the central CDW peak with temperature, namely the satellite peaks become closer to the central CDW peak with lower temperature. Since \textbf{q$_{domain}$} = \textbf{q$_{sat}$} – \textbf{q$_{cdw}$}, this indicates that the domain size grows with decreasing temperature as has been noted in previously.\cite{Thomson1994}

For each of the three temperatures, we extract the central CDW peak wavevector, \textbf{q$_{cdw}$}, using Gaussian fits to peaks in the FFTs, so as to determine the average CDW lattice periodicity and angle. Similar fits to satellite peaks at each temperature allow us to determine the average domain periodicity and angle relative to that of the atomic lattice. The results are summarized in Table \ref{table:1}. We find that the average CDW periodicity is essentially constant at 11.76 {\AA} over the temperature range, consistent with previous electron diffraction data \cite{Ishiguro1991}. The average angle of the CDW lattice with respect to the atomic lattice decreases with increasing temperature (from 12.6{\degree} to 10.5{\degree}) and has values consistent with both previous electron diffraction and STM measurements.\cite{Ishiguro1991, Thomson1994} The domain size/periodicity gets smaller (108 {\AA} at 203 K to 57 {\AA} at 354 K) as the temperature increases, consistent with previous STM data \cite{Thomson1994}. The satellite peak angle (equivalent to the domain lattice angle) with respect to that atomic lattice is close to that of the average CDW angle at 203 K (right above the C-CDW transition) and significantly evolves away from the average CDW angle with increasing temperature.

\begin{table}[h]
\centering
\renewcommand{\arraystretch}{1.5}
\begin{tabular}{||c || c | c | c||} 
 \hline
  & 203 K & 300 K & 354K \\ 
 \hline\hline
 CDW Periodicity ({\AA}) & 11.75 & 11.76 & 11.75 \\ 
\hline
 CDW Angle & +12.6{\degree}  & +11.8{\degree} & +10.5{\degree} \\ 
 \hline
 Domain Periodicity ({\AA}) & 108 & 73 & 57 \\ 
 \hline
 Domain Angle & +11.8{\degree} & +3.2{\degree} & -4.3{\degree} \\ 
 \hline
\end{tabular}
\caption{Table quantifying the average CDW and domain lattice structures at 203 K, 300 K, and 354 K. The average CDW quantities were determined using the central CDW peak, \textbf{q$_{cdw}$}, in each FFT. The average domain quantities were determined from \textbf{q$_{domain}$} = \textbf{q$_{sat}$} – \textbf{q$_{cdw}$}. The angle values are relative to the atomic lattice with '+' as counter clockwise and '-' as clockwise. Note, in the analysis above and in the simulations that follow, we mirror the 354 K FFT, given the two energetically equivalent CDW rotations relative to the atomic lattice, for consistent comparison with the 203 K and 300 K data and simulations.}
\label{table:1}
\end{table}

Following our previous work \cite{Singh2022}, we can use the wavevectors for the atomic peaks, central CDW peaks, and for the two most intense satellite peaks to simulate topographies for each of the three temperatures for further analysis. The simulations allow us to create clean, extended topographies so as to focus on the essential physics contained within the material without the influence of surface or subsurface defects which can otherwise complicate analysis. 

\begin{figure}[h]
\includegraphics[clip=true,width=\columnwidth]{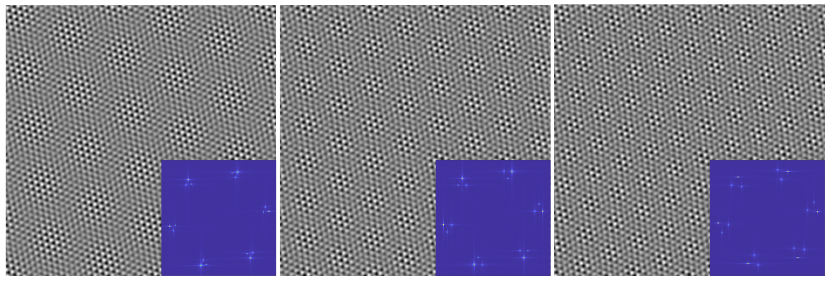}
\caption{500 \text{\AA} square computer simulated STM topographies for a) 203 K, b) 300 K, and c) 354 K. Topographies include the central CDW signal and two satellite peaks at the wavevectors found in the STM data (Figure \ref{fig:Fig2}). Insets are FFTs of the topography and illustrate these wavevectors. The relative intensities for the CDW and satellite peak signals used in each simulation were set by the experimentally-determined intensities in the 300 K FFT. This was done to eliminate differences in tip condition among the data sets and creates a consistency among the simulations. Important in this decision, in analyzing multiple topographies at each temperature, no temperature-dependent trend in peak intensities is found in the FFTs of STM topographies of the NC-CDW state.}
\label{fig:Fig3}
\end{figure}

Figure \ref{fig:Fig3} shows computer simulated topographies for the CDW lattice (central CDW peaks and two satellite peaks) at each of the three temperatures without the atomic lattice. Note that the simulated images are rotated such that the atomic lattice orientation is identical for each temperature simulation. In addition, in the case of the 354 K, we simulated the equivalent mirror image for consistency with the 203 K and 300 K simulations.

\begin{figure}[h]
\includegraphics[clip=true,width=\columnwidth]{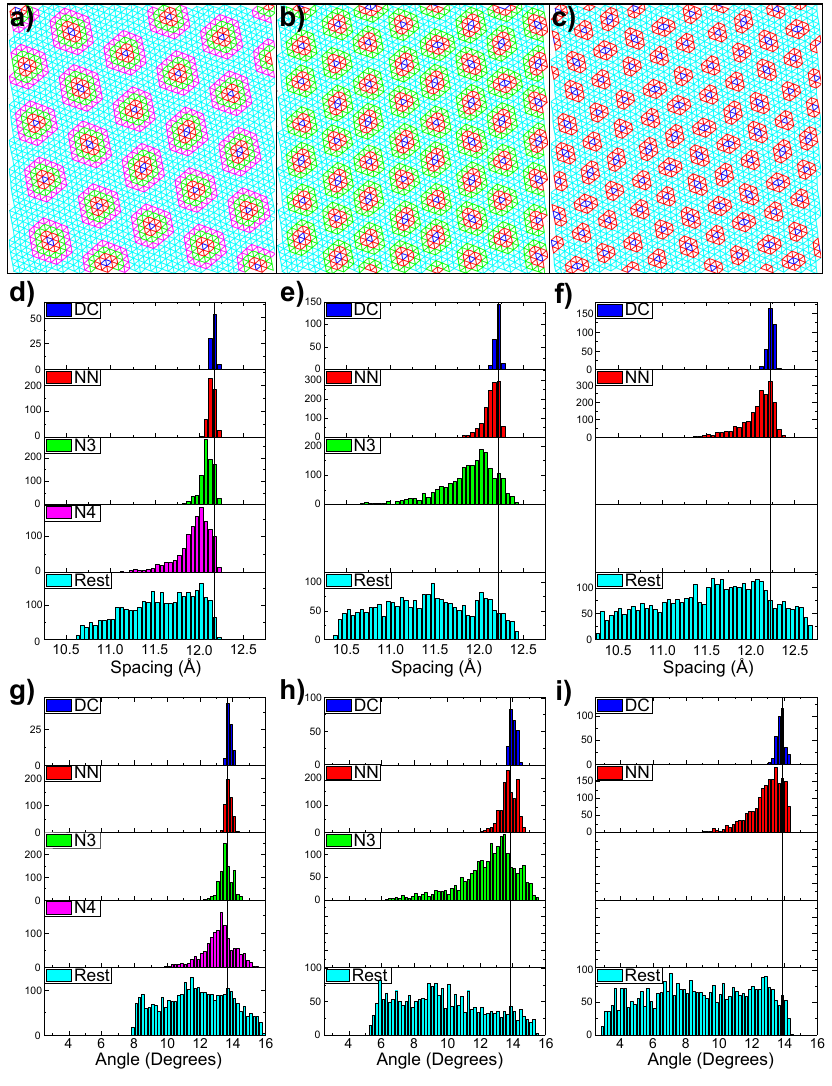}
\caption{a) 203 K, b) 300 K, and c) 354 K line segments connecting nearest neighbor CDW maxima in the images in Figure \ref{fig:Fig3}, color coded by connections among domain center (DC in blue), nearest neighbor CDW connectors to DC (NN in red), and so on from domain centers toward domain walls. d) 203 K, e) 300 K, and f) 354 K local CDW lattice spacing for each nearest neighbor line segment grouping. g) 203 K, h) 300 K, and i) 354 K local CDW lattice angle relative to the atomic lattice for each nearest neighbor line segment grouping.}
\label{fig:Fig4}
\end{figure}

We now examine the general evolution of the CDW lattice from domain center to domain wall. We use intensity thresholding to determine the 3 to 4 most intense CDW maxima within each domain for a simulated topography and use those to define the domain centers. We then draw line segments connecting these domain center CDW maxima and designate these line segments as 'DC' (seen in blue in Figures \ref{fig:Fig4}a-c). Next we find the CDW maxima which are nearest neighbors to those of the domain center. The line segments connecting these CDW maxima to those in the domain center and to each other we designate as NN (in red). We continue this process moving outward from the domain center connecting the next nearest neighbors to the domain center CDWs to the nearest neighbors and to each other (N3 in green) and so on. For a given temperature, domain-associated nearest neighbor line segments stop at the last color which includes no CDW maxima which would connect to two different domains (N4 for 203 K, N3 for 300 K, and NN for 354 K). The remaining CDW maxima comprise those mainly in domain walls. The nearest neighbor connections made from these remaining CDW maxima are colored in cyan and designated as 'Rest'. Together, using the drawn line segments, we can characterize domains from domain center to domain wall. The length of each segment gives the local CDW lattice spacing and the segment slope can be used to determine the local CDW lattice angle relative to that of the atomic lattice. Figure \ref{fig:Fig4} shows histograms which illustrate the progression of the nearest neighbor CDW distances and angles from domain center to domain walls. The statistics are summarized in Table \ref{table:2}.

At a given temperature, the average nearest neighbor CDW distance spacing and angle is largest at the domain center and the width of the distance/angle distributions are narrowest and close to the low-temperature C-CDW values of 12.1 \text{\AA} and 13.9\text{\degree} \cite{Tanda1985}. Progressing away from the domain center, the average nearest neighbor distances/angles become smaller and the distributions wider.

Comparing across temperatures, with lower temperatures, as domains get larger, the evolution in nearest neighbor distances/angles from domain center to domain wall is more gradual as compared to higher temperatures. Illustrating this, at 354 K, there is a $\sim$2.5 larger change in the average nearest neighbor spacing from DC to NN than at 203 K. Similarly, while there is a 0.9{\degree} change in angle from DC to NN at 354 K, there is only a 0.8{\degree} change in CDW angle from DC to N4 at 203 K. In addition, the width of each distribution for a given domain location (DC, NN, and N3) becomes narrower with lower temperature. In short, while there continues to be an evolution in the CDW lattice from domain center to domain wall at lower temperatures, the domain is becoming more uniform and closer to the CDW lattice found in the low-temperature C-CDW state.

\begin{table}[h]
\centering
\renewcommand{\arraystretch}{1.5}
\begin{tabular}{||c || c | c | c||} 
 \hline
 \rowcolor{lightgray} 
 Distance ({\AA}) & 203 K & 300 K & 354K \\ 
 \hline\hline
 DC & 12.16 $\pm$ 0.02 & 12.21 $\pm$ 0.03 & 12.23 $\pm$ 0.04 \\ 
\hline
 NN & 12.14 $\pm$ 0.04 & 12.14 $\pm$ 0.09 & 12.08 $\pm$ 0.18 \\ 
\hline
 N3 & 12.09 $\pm$ 0.07 & 11.88 $\pm$ 0.32 &  \\ 
\hline
 N4 & 11.94 $\pm$ 0.19 &  &  \\ 
\hline
 Rest & 11.55 $\pm$ 0.40 & 11.41 $\pm$ 0.54 & 11.56 $\pm$ 0.62 \\ 
\hline
 Domain Average & 11.78 $\pm$ 0.40 & 11.76 $\pm$ 0.50 & 11.77 $\pm$ 0.57 \\ 
\hline \hline

 \rowcolor{lightgray}
 Angle ({\degree}) & 203 K & 300 K & 354K \\ 
 \hline\hline
 DC & 13.8 $\pm$ 0.1 & 14.1 $\pm$ 0.2 & 13.8 $\pm$ 0.2 \\ 
\hline
 NN & 13.8 $\pm$ 0.2 & 13.8 $\pm$ 0.5 & 13.9 $\pm$ 1.0 \\ 
\hline
 N3 & 13.6 $\pm$ 0.4 & 12.5 $\pm$ 1.7 &  \\ 
\hline
 N4 & 13.1 $\pm$ 1.0 &  &  \\ 
\hline
 Rest & 11.7 $\pm$ 2.1 & 9.8 $\pm$ 2.7 & 8.8 $\pm$ 3.3 \\ 
\hline
 Domain Average & 12.5 $\pm$ 1.8 & 11.7 $\pm$ 2.6 & 10.5 $\pm$ 3.3\\ 
\hline 

\end{tabular}
\caption{Table quantifying the CDW nearest neighbor spacing and angles (relative to the atomic lattice) for each line segment grouping in Figure \ref{fig:Fig4}.}
\label{table:2}
\end{table}

\begin{figure}[b]
\includegraphics[clip=true,width=\columnwidth]{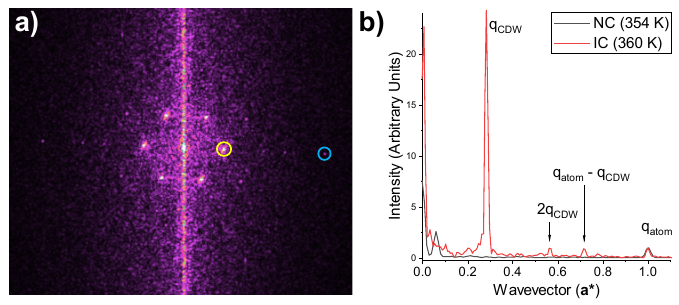}
\caption{a) FFT of STM topography taken at 360 K in the IC-CDW state. One atomic peak is circled in blue. One CDW peak is circled in yellow. There are no satellite peaks surrounding the CDW peak in the IC-CDW state.  b) In red, a linecut from FFT center through the atomic peak circled in a) which also passes through the circled CDW peak. This linecut, taken in the IC-CDW phase, shows a clear CDW peak at 0.28 \textbf{a*}. In black, a comparable linecut from the 354 K FFT (NC-CDW state at a temperature just below the IC-CDW transition), the center of which appears in Figure \ref{fig:Fig2}c, through an atomic peak shows no corresponding peak CDW-relate peaks.
}
\label{fig:Fig5}
\end{figure}

At higher temperatures, the domains progressively become smaller, and the fraction of the sample which is comprised of the domain walls increases. As a rough estimate of this fraction, we can compare the percentage of nearest neighbor connectors designated as ‘Rest’ at each of the temperatures: 53\% at 203 K and 63$\%$ at 354 K. The domain walls separating domains have been described as discommensurations or an incommensurate wall network.\cite{Ma2016, Shao2016} Since a domained NC-CDW state emerges from the IC-CDW with cooling through the $\sim$355 K transition, it is natural to investigate whether there is evidence for the higher-temperature IC-CDW state within the NC-CDW state, possibly within the domain walls. Figure \ref{fig:Fig5} shows the FFT of an STM topography of the IC-CDW state we acquired at 360 K. The FFT shows that, in the IC-CDW state, \textbf{q$_{CDW}$} is colinear with the atomic lattice and a linecut in the direction of \textbf{a*} reveals a CDW peak at (\textbf{q} = 0.28 \textbf{a*}), consistent with what has been previously reported by x-ray diffraction measurements \cite{Nakanishi1977}. A similar linecut through the FFT of NC-CDW state at 354 K (from Figure \ref{fig:Fig2}c) in the direction of \textbf{a*}, shows no peak near \textbf{q} = 0.28 \textbf{a*} and no evidence for the presence of the IC-CDW state within the NC-CDW state. Instead of a smooth evolution from the NC-CDW state to the IC-CDW state, there is a sudden shift in \textbf{q$_{CDW}$} in passing through the transition.

\subsection{Chirality}
\begin{figure}[h]
\includegraphics[clip=true]{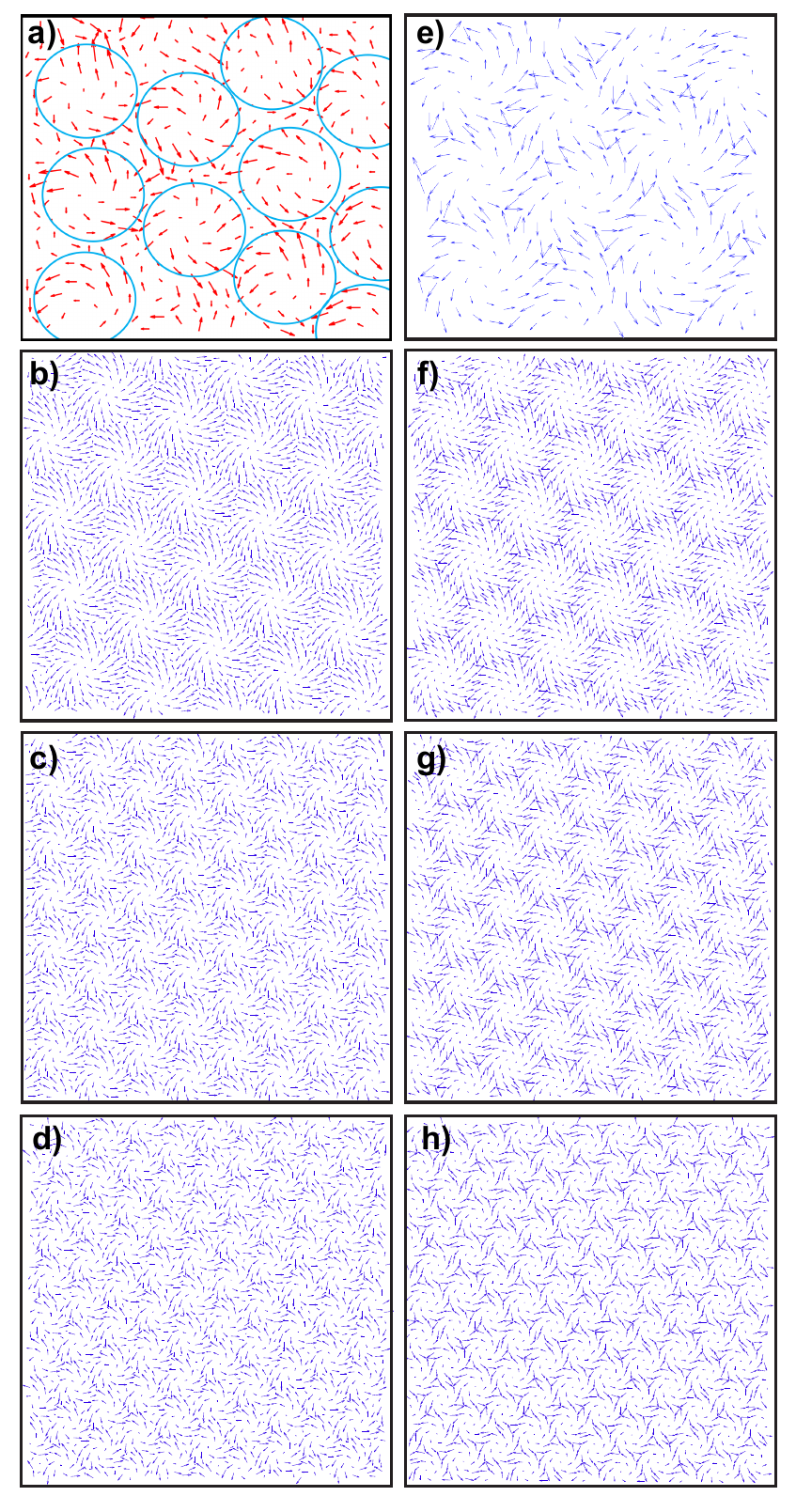}
\caption{a) CDW$_{ave}$ - CDW$_{ltc}$ displacement field for the STM topography in Figure \ref{fig:Fig1}. Arrows point from CDW$_{ave}$ maxima to CDW$_{ltc}$ maxima and are proportional to the displacement. Circled spiral domains match the visible domains in the original topography. b) 203 K, c) 300 K, d) 354 K CDW$_{ave}$ - CDW$_{ltc}$ displacement fields for the simulated topographies of Figure \ref{fig:Fig3}. e) CDW$_{ltc}$ - atom displacement field for the STM topography in Figure \ref{fig:Fig1}. f) 203 K, g) 300 K, h) 354 K displacement fields for CDW$_{ave}$ to nearest atom. Figures a) and e) are from \cite{Singh2022}.
}
\label{fig:Fig6}
\end{figure}

Previously, an intradomain chirality characterizing the NC-CDW state at 300 K was uncovered by examining the displacement field created by examining the shifts of the actual CDW lattice (CDW$_{ltc}$) relative to the average CDW lattice (CDW$_{ave}$).\cite{Singh2022} The CDW$_{ave}$ lattice is the result of Fourier filtering an STM topography such that only the signals from the central CDW peaks (one of which is designated by a white circle in Figure \ref{fig:Fig1}b) are present. This results in a perfect hexagonal lattice incommensurate with the atomic lattice. CDW$_{ltc}$, which is the fully observed CDW lattice with domaining in the topography, can be isolated from an STM topography by Fourier filtering the central CDW peak and surrounding satellite peaks. CDW$_{ltc}$ peaks are shifted slightly from CDW$_{ave}$ peaks, by an average of 0.72 \text{\AA} and a maximum displacement of 1.70 \text{\AA}. Arrows drawn from each CDW$_{ave}$ peak location to the nearest CDW$_{ltc}$ peak location with a length proportional to the separation creates the displacement field seen in Figure \ref{fig:Fig6}a. The intradomain chirality is obvious by eye with a counterclockwise spiral of arrows within each domain. Displacements are smallest at domain centers and continue to increase away from domain centers to domain walls. Computer simulated STM topographies using the signals associated with the central CDW peak and two most intense satellite peaks well recreate this displacement field.\cite{Singh2022}

We build on this previous work and extend this to 203 K and 354 K using STM-data-based CDW computer simulated topographies to understand the evolution of the displacement fields across the NC-CDW state of 1T-TaS$_{2}$. Figures \ref{fig:Fig6}b-d show the displacement fields for the three temperatures 203K, 300K, and 354 K. In each, the spiral nature of the displacement field yielding the intradomain chirality is obvious. However, despite the different domain sizes at each of the three temperatures, the average displacement remains essentially constant with average displacements of 0.78 $\pm$ 0.31 \text{\AA}, 0.75 $\pm$ 0.30 \text{\AA}, and 0.75 $\pm$ 0.31 \text{\AA} at 203 K, 300 K, and 354 K respectively.

Figures \ref{fig:Fig6}e-h show the displacement field resulting from connecting CDW$_{ave}$ maxima to the nearest atomic lattice site. Like those in Figures \ref{fig:Fig6}a-d, these displacement fields show a spiral-like domain structure with displacements smallest at the center of a domain. The displacement fields in \ref{fig:Fig6}e-h represent the displacements of the CDW$_{ave}$ in the limit where the CDW lattice has a strong coupling to the atomic lattice with no energy cost for deformation. The striking similarity of these displacement fields strongly suggest that a strong coupling of the CDW lattice to the atomic lattice drives the intradomain chirality. However, there are some differences between the displacement fields. For example, we find that, at a given temperature, the domain size and orientation from the displacement field for CDW$_{ave}$ to CDW$_{ltc}$ is slightly different from that of CDW$_{ave}$ to nearest atomic sites (Table \ref{table:3}). These differences point to additional physics at play. As proposed previously \cite{Singh2022}, there likely is a competition between CDW lattice coupling and energy gained from the electronic structure. Both CDW-lattice coupling and the electronic structure likely evolve with temperature in ways which require more understanding. Further insight into this competition would benefit from theoretical work, such as using a McMillan-type Landau framework \cite{McMillan1975,McMillan1976} which specifically includes CDW-atomic lattice coupling.

\begin{table}[h]
\centering
\renewcommand{\arraystretch}{1.5}
\begin{tabular}{||c || c | c ||} 
 \hline
  & Domain Size Comparison & Domain Angle \\
  & (Percent Difference) & \\
 \hline\hline
203 K & 9.6 & -2.4{\degree} \\ 
\hline
300 K & 2.1 & -1.5{\degree} \\ 
 \hline
354 K & 14.7 & +3.2{\degree} \\
 \hline
\end{tabular}
\caption{Table comparing the domain size and orientation for the CDW$_{ave}$ - CDW$_{ltc}$ displacement field as compared to that of the CDW$_{ave}$ to nearest atomic site displacement field.
For each temperature, the CDW$_{ave}$ - CDW$_{ltc}$ domain size is larger. 'Domain Angle' is the angle between the hexagonal domain lattice produced by the CDW$_{ave}$ - CDW$_{ltc}$ displacement field and the CDW$_{ave}$ - atomic displacement field. ‘-’/’+’ means the domains of CDW$_{ave}$ - CDW$_{ltc}$ displacement field are rotated ‘clockwise’/’counter clockwise’ relative to the domains of the other displacement field.
}
\label{table:3}
\end{table}

\begin{figure}[h]
\includegraphics[clip=true,width=45 mm]{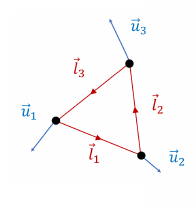}
\caption{We determine the curl of the displacement field by calculating the circulation of the displacement field, \textbf{u}, for each set of three CDW$_{ave}$ maxima and dividing by the area of the triangle. Here we calculate the circulation counter clockwise: \textbf{u$_{1}$}, \textbf{u$_{2}$}, and \textbf{u$_{3}$} are the three displacement field vectors associated with each of the three CDW$_{ave}$ maxima and \textbf{l$_{1}$}, \textbf{l$_{2}$}, and \textbf{l$_{3}$} are the path directions and lengths connecting these maxima. The circulation is calculated using Equation 2 and is positive/negative for counterclockwise/clockwise circulation.
}
\label{fig:Fig7}
\end{figure}

We use the curl of the displacement field (twice the rotation vector, the antisymmetric component of the displacement gradient tensor) as a way to visualize chirality and quantify the displacement fields of Figure \ref{fig:Fig6} and their evolution with temperature. We calculate the curl of the displacement field, \textbf{u}, following Stoke's theorem: 
\begin{equation}
\int_S\mbox{curl}\,{\bf u}\cdot{\bf dA}=\oint_C{\bf u}\cdot{\bf dl}
\end{equation}
We calculate the curl by calculating the circulation of the displacement field for each group of three CDW lattice sites and dividing by the area of the triangle formed by the three sites as seen in Figure \ref{fig:Fig7}. Given the discrete nature of the displacement field we calculate the circulation for the triangle in Figure \ref{fig:Fig7} using:
\begin{equation}
\oint_C{\bf u}\cdot{\bf dl}\ {\sim} \frac{1}{2}\left[{\bf u_1}\cdot\left({\bf l_1} + {\bf l_3}\right) + {\bf u_2}\cdot\left({\bf l_1} + {\bf l_2}\right) + {\bf u_3}\cdot\left({\bf l_2} + {\bf l_3}\right)\right]
\end{equation}

\begin{figure}[h]
\includegraphics[clip=true]{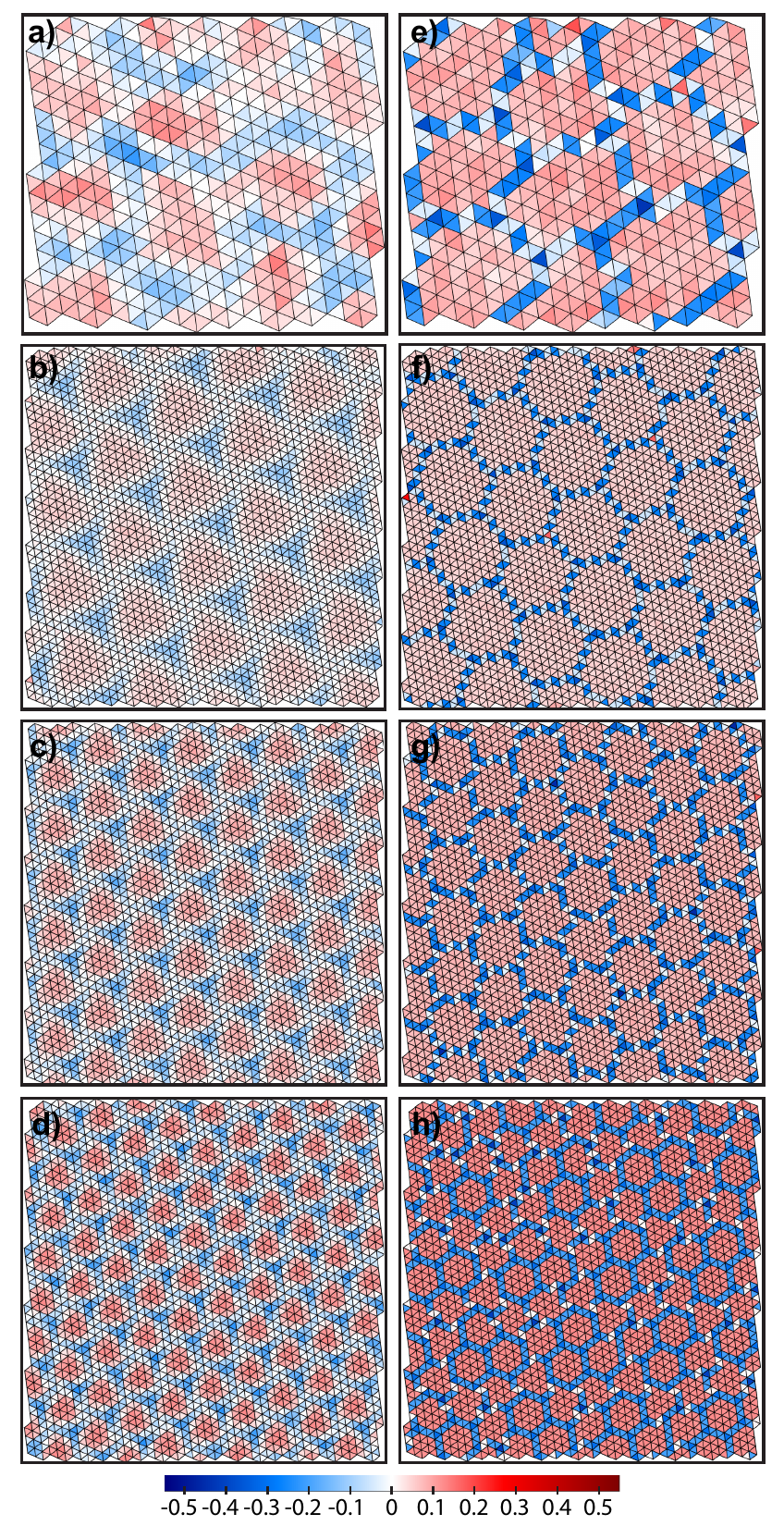}
\caption{Plots of the curl of the displacement fields seen in Figure \ref{fig:Fig6}. a)-d) Plots of the curl of the CDW$_{ave}$ - CDW$_{ltc}$ displacement fields for a) STM-acquired topography at 300 K (a) and for the extended simulated topographies b) 203 K, c) 300 K, d) 354 K. e)-h) Plots of the curl of the CDW$_{ave}$ - atom displacement fields for STM-acquired topography at 300 K (e) and for the extended simulated topographies f) 203 K, g) 300 K, h) 354 K.
}
\label{fig:Fig8}
\end{figure}

Figure \ref{fig:Fig8}a-d shows the curl calculations for the CDW$_{ave}$ - CDW$_{ltc}$ displacement fields seen in Figure \ref{fig:Fig6}a-d. At a given temperature, within a domain, the curl has one sign (we choose positive) and is roughly constant until the domain edge. Within the domain walls, the curl is negative and is largest at the point of convergence of three domains. However, the domains are not C6 rotationally symmetric as one might expect for a hexagonal domain pattern. Rather, the domains have a more triangular shape more in line with C3 rotational symmetry as is most obvious in the 203 K curl diagram. Consistent with this, within the domain wall at the locations where three domains converge of which there are six, three of the locations show a larger magnitude curl (darker blue) than the other three. The average curl across the topographic region is zero. While the domains have one chirality, the domain walls have an opposite chirality. And, while there is local chirality, this does not lead to large-scale chirality (aside from the rotation of the CDW lattice with respect to the atomic lattice) which may explain why only recently this intradomain chirality in 1T-Ta$_{2}$ was uncovered.

Examining the curl of the CDW$_{ave}$ - CDW$_{ltc}$ displacement fields with temperature, the magnitude of the curl within a domain, as well as within domain walls, increases with increasing temperature. However, summing the curls for each triangle within a domain together, we find: 2.8 (203 K), 2.1 (300 K), 1.8 (354 K). This indicates that while the curl within a domain decreases with decreasing temperature, the circulation for the entire domain, $\int_{Domain}\mbox{curl}\,{\bf u}\cdot{\bf dA}=\oint_{Domain}{\bf u}\cdot{\bf dl}$, increases.

For comparison, we calculate the curl of the CDW$_{ave}$ – atomic displacement field (Figures \ref{fig:Fig8}e-h). Similar to the curls in Figures \ref{fig:Fig8}a-d, the curls of these displacement fields are roughly constant and positive within a domain and have the opposite sign in the domain walls. The magnitude of the curl within a domain, as well as within domain walls, increases with increasing temperature, also consistent with the Figures \ref{fig:Fig8}a-d. If we sum the curls for each triangle within a domain, we find: 5.0 (203 K), 3.9 (300 K), 2.7 (354 K). Again we find the overall circulation for a domain increases with decreasing temperature.

Differing from the curl of the CDW$_{ave}$ – CDW$_{ltc}$ displacement fields, the curls for the CDW$_{ave}$ – atomic displacement fields show an abrupt change in the curl sign at the domain walls, the domain walls are thinner, and the domains appear hexagonally-shaped with a roughly C6 rotational symmetry. These differences further emphasize that physics beyond a strong coupling of the CDW to the lattice drives the CDW$_{ltc}$ properties in 1T-TaS$_{2}$.

\subsection{Conclusions}
In our sub-angstrom characterization of domains across the NC-CDW state of 1T-Ta$_{2}$, we find that, at any temperature, there is an evolution in the nearest neighbor CDW spacing and angle relative to the lattice from domain center to domain wall. At the center of a domain, the CDW spacing and local angle are close to that of the low temperature C-CDW state. But, as one progresses from the center to the domain wall, the average spacing and local angles decrease, and there is a wider variation in the spacing/angle values.

As the temperature decreases and the domains become larger, there are smaller differences in the distances/angles for DC, NN, N3 line segments. In essence, the domains are becoming more uniform with a higher fraction of distances/angles close to that of the low-temperature C-CDW state. At 354 K, just below the transition to the IC-CDW state, we find no evidence for the presence of the IC-CDW state. Instead, there is a sudden change of \textbf{q$_{cdw}$} through the first order phase transition from the NC-CDW state at 354 K to the IC-CDW state, most noticeable by the sudden rotation of the wavevector.

The NC-CDW state is characterized by an intradomain chirality. Characterizing the rotation vector for the CDW$_{ave}$ – CDW$_{ltc}$ displacement field shows that the curl is roughly constant within a domain center and has the opposite sign in the domain walls. Whereas the average displacements for the CDW$_{ave}$ – CDW$_{ltc}$ displacement field is unchanged with temperature, the curl magnitude within the domains/domain walls increases with temperature. This trend is similar to that of the CDW$_{ave}$ – atomic displacement field which is the displacement field which would be found if one were to start with an incommensurate CDW lattice and suddenly turn on coupling to the atomic lattice with no energy cost for distortions. The similarity in trends provides evidence that the formation of chiral domains is strongly related to the coupling of the CDW to the lattice. However, the differences in the displacement fields and curls point to a possible competition between coupling of the CDW to the atomic lattice and energy gained from the electronic structure.

Within the IC-CDW state, the curl is zero since CDW$_{ave}$ and CDW$_{ltc}$ are the same; there are no satellite peaks. The sudden onset of the large curl below $\sim$355 K accompanies the 1st order phase transition. As the temperature decreases and domain size increases, the curl magnitudes within the domains/domain walls decrease. This indicates a generally smoother evolution towards the C-CDW state in which the curl is 0 because again CDW$_{ave}$ and CDW$_{ltc}$ are the same within the C-CDW state. However, despite this smoother evolution for the curl, the NC-CDW to C-CDW phase transition is first order \cite{Wang2019, Kratochvilova2017} Future work which includes STM topographic measurements taken at small temperature increments around both NC-CDW to IC-CDW and NC-CDW to IC-CDW phase transitions will give stronger insight into the atomic-scale evolution of the CDW state through the transitions.

\subsection{Methods}
STM measurements were made using an RHK PanScan Freedom STM at 203 K, 300 K, 354 K, and 360 K. We used a tungesten tip, chemically-etched, cleaned, and sharpened via electron bombardment. The high-quality 1T-TaS$_2$ crystals used in this study were purchased from HQ Graphene. Samples were cleaved at $\sim$$10^{-8}$ Torr. 

\subsection{Acknowledgements}
We appreciate our informative conversations with Jasper van Wezel. This work was supported by the National Science Foundation under Grant No. DMR-1904918.


\bibliography{TaS2.bib}  

\end{document}